\documentclass{pnastwo}

\usepackage{pnastwoF}
\usepackage{amssymb,amsfonts,amsmath}
\usepackage[pdftex]{graphicx}
\usepackage{bm}
\usepackage{color}

\begin{document}

\title{Unraveling quantum mechanical effects in water using isotopic fractionation}

\author{Thomas E. Markland\affil{1}{Department of Chemistry, Stanford University, Stanford, California 94305-5080}
\and B. J. Berne\affil{2}{Department of Chemistry, Columbia University, New York, New York 10027}}

\maketitle

\begin{article}

\begin{abstract}
  When two phases of water are at equilibrium, the ratio of hydrogen
  isotopes in each is slightly altered due to their different phase
  affinities. This isotopic fractionation process can be utilized to
  analyze water's movement in the world's climate. Here we show that
  equilibrium fractionation ratios, an entirely quantum mechanical
  property, also provide a sensitive probe to assess the magnitude of
  nuclear quantum fluctuations in water. By comparing the predictions
  of a series of water models, we show that those describing the OH
  chemical bond as rigid or harmonic greatly over-predict the
  magnitude of isotope fractionation. Models that account for
  anharmonicity in this coordinate are shown to provide much more
  accurate results due to their ability to give partial cancellation
  between inter and intra-molecular quantum effects. These results
  give evidence of the existence of competing quantum effects in water
  and allow us to identify how this cancellation varies across a wide
  range of temperatures. In addition, this work demonstrates that
  simulation can provide accurate predictions and insights into
  hydrogen fractionation.
\end{abstract}

\keywords{water | quantum effect | isotope | fractionation}

\dropcap{W}ater within earth's atmosphere is naturally composed of the stable
hydrogen isotopes Hydrogen (H) and Deuterium (D). During cycles of
evaporation, condensation and precipitation, these isotopes naturally
undergo partial separation due to their differing masses thereby
leading to different H/D ratios in the two phases. This process of
fractionation has a number of fortuitous consequences which are
utilized in hydrology and geology. For instance, by comparing the
ratio of H to D, one can estimate the origins of a water sample, the
temperature at which it was formed, and the altitude at which
precipitation occurred \cite{Hoefs1997,Worden2007}. Equilibrium
fractionation, where the two phases are allowed to equilibrate their
H/D ratio, is entirely a consequence of the effects of quantum
mechanical fluctuations on water's hydrogen bond network.
Quantum mechanical effects such as zero-point energy and
  tunneling are larger for H due to its lower mass.

Despite numerous studies, the extent to which quantum fluctuations
affect water's structure and dynamics remains a subject of
considerable debate. It has long been appreciated that one effect of
quantum fluctuations in water is the disruption of hydrogen bonding,
leading to de-structuring of the liquid and faster dynamics
\cite{Kuharski1984,Wallqvist1985,Paesani2006,Miller2005}. However,
more recent work has suggested that a competing quantum effect may
exist in water \cite{Chen2003,Habershon2009}, namely that the quantum kinetic energy
 in the OH covalent bond allows it to stretch and form
shorter and stronger hydrogen bonds, which partially cancels the
disruptive effect. This hydrogen bond strengthening has only been
recently appreciated, as many original studies drew their conclusions
based on models with rigid or harmonic bonds which are unable to
describe this behavior. The degree of quantum effect cancellation
depends sensitively on the anharmonicity of the OH stretch and the
temperature. These parameters tune the balance between the lower
frequency hydrogen bonding disruption, which will dominate at lower
temperatures, and the higher frequency hydrogen bond strengthening
effect, which will dominate at higher temperatures when rotations
become essentially classical.

If such a large degree of cancellation existed at ambient temperature,
it would be highly fortuitous both in terms of the biological effects
of heavy water, which is only mildy toxic to humans
\cite{Kushner1999}, as well as the ability to use heavy solvents in
2D-IR and NMR spectroscopies, where deuteration is assumed not to
dramatically alter the structure or dynamics observed. However, the
size of this cancellation remains elusive since empirical quantum
models of water are typically fit to reproduce its properties when
used in path integral simulations and the two {\em ab initio} path
integral studies performed have not produced a consistent picture
\cite{Chen2003,Morrone2008}. In addition, many of these simulation
studies compare the properties of water to those of its classical
counterpart, but ``classical'' water is physically unrealizable even
at relatively high temperatures, since water still has significant
quantum effects present in its vibrations.

In this paper, we use equilibrium fractionation ratios as a sensitive
probe to assess the magnitude of quantum mechanical effects in
water. Fractionation ratios can be directly related to quantum kinetic
energy differences between H and D in liquid water and its vapor
and can be calculated exactly for a given water
potential energy model using path integral simulations. The large
number of accurate experimental measurements of these ratios
allows for sensitive comparisons of theory and
experiment over a wide range of temperatures \cite{Horita1994}. In the
present work, we show what features are needed in a water model to
accurately predict these ratios by decomposing the contributions to
the free energy difference leading to fractionation. This in turn
leads to a simple explanation of the inversion of the fractionation
ratios seen experimentally at high temperatures, where D is favored
over H in the vapor phase\cite{Horita1994}.

\section{Calculating Fractionation Ratios}

\begin{figure}[h]
\centerline{\includegraphics[width=.4\textwidth]{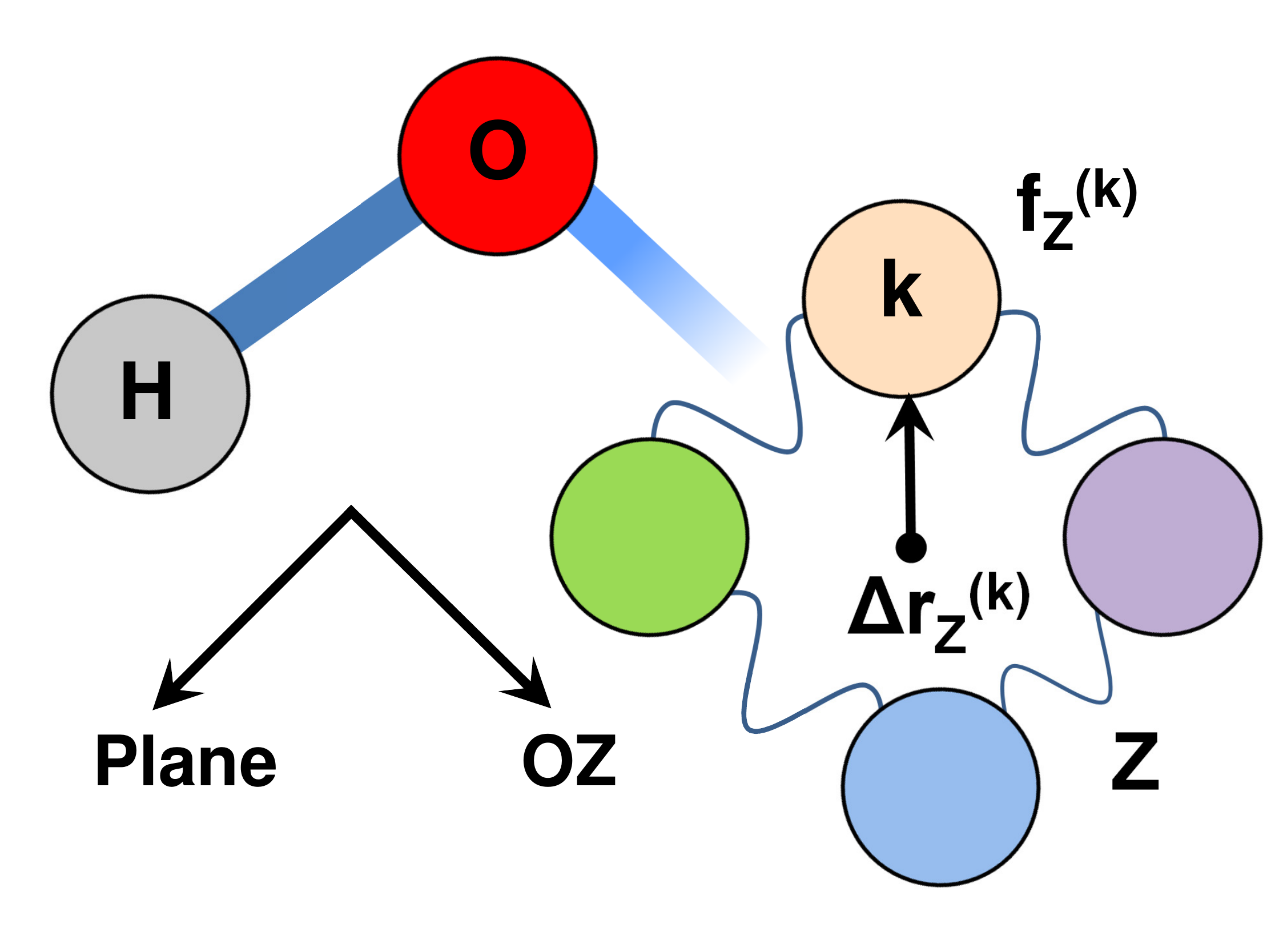}}
  \caption{Graphical representation of the vectors introduced in the text for a water molecule HOZ 
  where Z is a hydrogen isotope of mass, $m_{Z}$. For clarity the H and O atoms are shown as 
  single atoms whereas in the PIMD formalism they are also mapped onto ring polymers.}
  \label{fig:waterbeads}
\end{figure}

The liquid-vapor fractionation ratio, $\alpha_{l-v}$, is defined as
\begin{equation}
\alpha_{l-v} = \frac{(x_{D,l}/x_{H,l})}{(x_{D,v}/x_{H,v})} = e^{-\Delta A/k_{B}T},
\end{equation}
where $x_{Z}$ is the mole fraction of isotope $Z$, $l$ denotes the
liquid phase, and $v$ denotes the vapor phase. In the second equality,
$\Delta A$ is the Helmholtz free energy corresponding to the process
\begin{equation}
H_{2}O_{(l)} + HOD_{(v)} \rightleftharpoons H_{2}O_{(v)} + HOD_{(l)}.
\end{equation}
In this work we consider the dilute deuterium (D) limit which reflects
the situation found in the Earth's atmosphere where it is 6000 times
less common than H. In this limit, we consider the free energy of
exchanging a single D atom in a vapor water molecule with an H atom in
a liquid water molecule, with all other molecules being H$_{2}$O. The
free energy difference can be calculated from the thermodynamic
integration expression \cite{Vanicek2007}
\begin{equation}
  \Delta A = \int^{m_{D}}_{m_{H}}  dm_{Z} \left( \frac{\left<K_{l}(m_{Z})\right>-\left<K_{v}(m_{Z})\right>}{m_{Z}} \right),
\label{eq:deltaA}
\end{equation}
where $\left<K_{l}(m_{Z})\right>$ and $\left<K_{v}(m_{Z})\right>$ are
the kinetic energy expectation values for a hydrogen isotope of mass
$m_{Z}$ in a water molecule HO$Z$ in the liquid and vapor phases
respectively. The kinetic energy can be calculated exactly for a given
potential energy model of water using a path integral molecular
dynamics (PIMD) simulation. These simulations exploit the exact
isomorphism between a system of quantum mechanical particles and that
of a set of classical ring polymers in which the spread of a polymer
is directly related to that quantum particle's position
uncertainty\cite{Parrinello1984}. The kinetic energy for the particle
$Z$ in the molecule HO$Z$ can be calculated from these simulations
using the centroid virial estimator \cite{Herman1982,berne3_1989}
\begin{equation}
  K =  \frac{3 k_{B} T}{2} - \frac{1}{2n} \sum_{k=1}^{n} \Delta {\bf r}_{Z}^{(k)} \cdot {\bf f}^{(k)}_{Z},
\label{eq:kinetic}
\end{equation}
where T is the temperature and k$_{B}$ is the Boltzmann
constant. Here, $\Delta {\bf r}_{Z}^{(k)}$ is vector from the $k$th
bead representing particle $Z$ to the center of the ring polymer and
${\bf f}^{(k)}_{Z}$ is the force on that bead as shown in
Fig. \ref{fig:waterbeads}. The first term in this expression is the
classical kinetic energy and is independent of the surrounding
environment, thus it is identical in the vapor and liquid phase. The
second term is the kinetic energy associated with confinement of a
quantum particle. This confinement depends on the forces exerted on
the particle by the surrounding molecules. Equilibrium fractionation
is thus an entirely quantum mechanical phenomenon.

PIMD simulations were performed using 1000 water molecules with $n$=32
ring polymer beads used for the imaginary time
discretization. Previously described evolution and thermostatting
procedures were used \cite{Ceriotti2010}. The computational cost of
these calculations was reduced by using the ring polymer contraction
technique with a cut-off of 5\AA, which for this system size leads to
more than an order of magnitude speed-up compared to a
standard PIMD implementation \cite{Markland2008,Markland2008b}. The
integral in Eq. \ref{eq:deltaA} was performed using the midpoint
rule with 11 masses $m_{Z}$ evenly spaced between $m_{H}$ and
$m_{D}$. Calculations were performed at the experimental coexistence
densities\cite{Wagner2002}.

To model the interactions within and between water molecules, we used
the q-SPCFw \cite{Paesani2006} and q-TIP4P/F \cite{Habershon2009}
models which have previously been shown to accurately reproduce many
of water's properties in PIMD simulations of liquid water. Both models
are flexible, use point charges, and have a harmonic description of
the bending mode. However, while q-SPCFw uses a purely harmonic
description of the OH stretch
\begin{equation}
V_{\rm OH}(r) = \frac{1}{2}k_r(r-r_{\rm eq})^2,
\label{eq:harm}
\end{equation}
q-TIP4P/F contains anharmonicity by modeling the stretch as a Morse
expansion truncated at fourth order
\begin{equation}
V_{\rm OH}(r) =  D_r \left[ \alpha_r^2(r-r_{\rm eq})^2-\alpha_r^3(r-r_{\rm eq})^3 
              +  \frac{7}{12}\alpha_r^4(r-r_{\rm eq})^4 \right].
\label{eq:aharm}
\end{equation}
Here $r$ is the distance between the oxygen and hydrogen atom and the
parameters are given in Refs. \cite{Paesani2006} and
\cite{Habershon2009}. The anharmonicity in the q-TIP4P/F model makes
the observed quantum mechanical effects much smaller than previously
predicted from harmonic or rigid models and gave rise to the idea of
``competing quantum effects'' in water\cite{Habershon2009}. 

  Both models have previously been shown to
  accurately reproduce many of water's properties in PIMD simulations
  of liquid water. Due to their simple potential form such models are generally
  less transferable to other phases than more sophisticated
  polarizable or {\em ab initio} descriptions. However partially adiabatic centroid molecular dynamics simulations \cite{Hone2006} have 
  shown that the anharmonic stretch (Eq. \ref{eq:aharm}) allows reasonable agreement to be obtained in the
  observed frequency shifts in the infrared spectrum in going from
  liquid to gaseous water as well as from pure light to pure heavy water \cite{Habershon2009,Habershon2008,Paesani2010}. 
  These models were chosen to assess the importance of anharmonicity in the OH
  stretch using a ``zeroth order'' description of liquid water. Hence,
  as we show below, they offer a straightforward way to assess
  competing quantum effects in water.

\section{Results and Discussion}

\begin{figure}[h]
\centerline{\includegraphics[width=.4\textwidth]{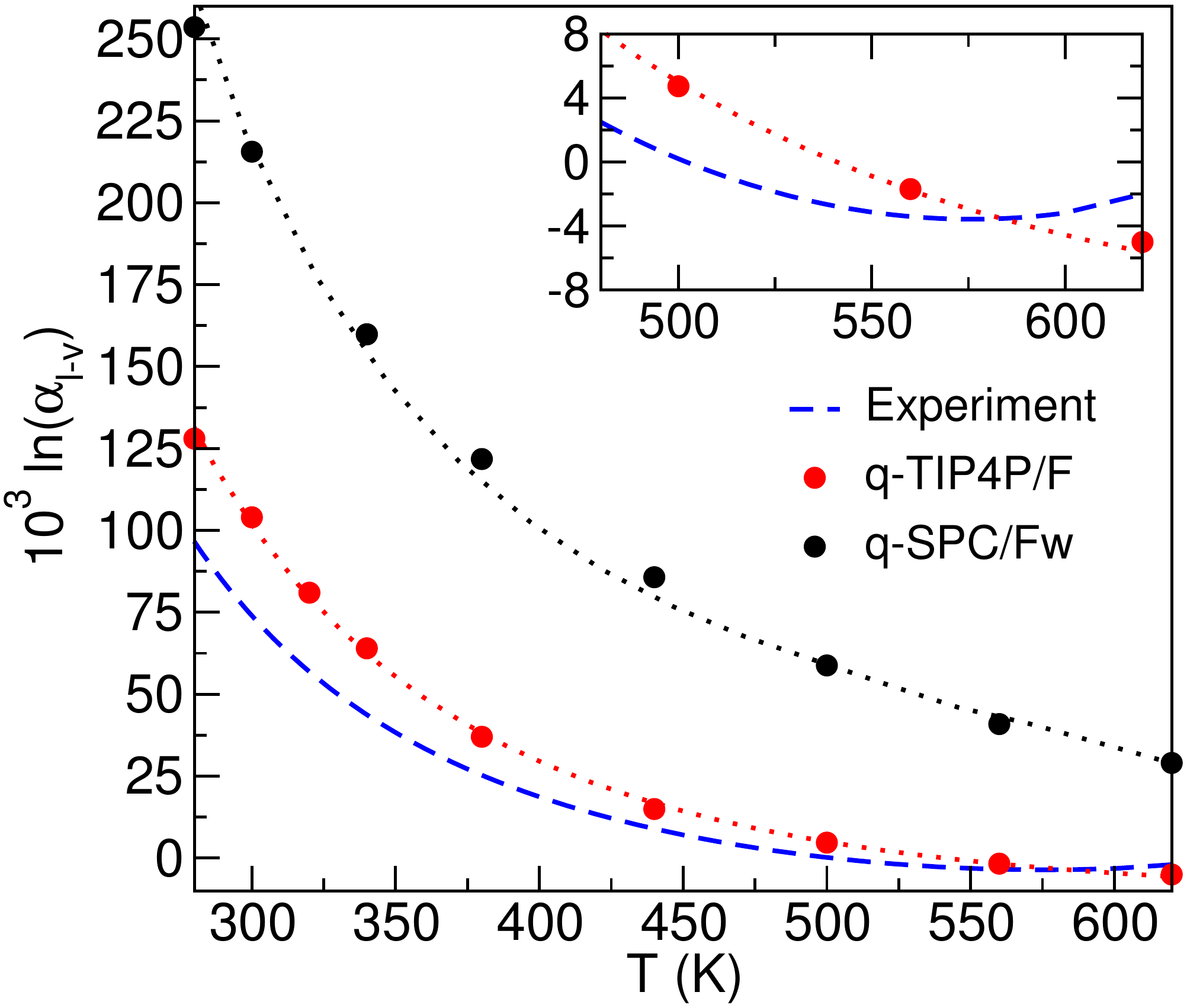}}
  \caption{Fractionation factors as a function of temperature for the
    q-SPCFw and q-TIP4P/F water models which describe the OH bond
    stretch using harmonic and anharmonic functions respectively. The
    experimental results are taken from Ref. \cite{Horita1994}. The inset shows the high temperature
    region where both experiment and the q-TIP4P/F model show an inversion in the fractionation where D is preferentially found in
    the vapor phase.  The curves through the calculated points are simply guides for the eye.}
  \label{fig:graph}
\end{figure}

Figure \ref{fig:graph} shows the fractionation factors calculated from our
PIMD simulations compared to the experimental data of
Ref. \cite{Horita1994}. For consistency with the experimental
data, we plot $10^3 \ln \alpha_{l-v}$ which is simply $-10^3 \Delta A/
k_{B} T$. Since $\Delta A/ k_{B} T$ is generally small, $e^{-\Delta A/
  k_{B} T} \simeq 1 - \Delta A/ k_{B} T$ thus an experimental value at
280 K of 100 corresponds to 10\% more D residing in the liquid than
the vapor. Above 500 K, the experimental data shows a well
characterized region of inverse fractionation where D becomes more favored in the
vapor than in the liquid, as shown in the inset of Fig. \ref{fig:graph}.

Turning to the simulated data, we observe that the harmonic q-SPC/Fw
model over-predicts the magnitude of fractionation at 300 K by a
factor of 3, and does not fall to the value observed
experimentally at 300 K until the temperture is raised to 450 K. In
contrast, the q-TIP4P/F model is in error by only 25\% at the lowest
temperature and approaches the experimental values more closely at
higher temperatures. It also correctly shows inverse fractionation
above 540 K. A previous study using the rigid SPC/E model and
$\hbar^{2}$ perturbation theory, found a H/D fractionation of 450 at 300 K
which is $\sim$5 times higher than that seen experimentally
\cite{Chialvo2009}. However, it is not clear whether this was
purely due to the use of a rigid model or whether the approximate
$\hbar^{2}$ perturbation technique used to obtain the fractionation
ratios was also at fault.

While Fig.\ref{fig:graph} demonstrates that the q-TIP4P/F model
provides much better agreement with the experimental data than
q-SPCFw, it is not immediately clear what aspect of the
parameterization causes this. To better understand the origins of this
effect, we constructed two models which we denote Aq-SPC/Fw and
Hq-TIP4P/F. In the former, the q-SPC/Fw water model has its harmonic
OH stretch replaced by a fourth order Morse expansion using the
parameters of q-TIP4P/F; in the latter, the Morse potential of
q-TIP4P/F is truncated at the harmonic term (see Eqs. \ref{eq:harm}
and \ref{eq:aharm}). The anharmonic variant of q-SPC/Fw gives results
as good as q-TIP4P/F, while the harmonic version of q-TIP4P/F fares as
poorly as q-SPCFw. In other words, the accurate prediction of the
fractionation ratios in liquid water is tied to the anharmonicity in
the OH direction and is rather insensitive to the other
parameters. This sheds light on a previous study where a sophisticated
rigid polarizable model gave identical predictions for H/D
fractionation to the simple fixed-charge rigid SPC/E model,
i.e. varying the intermolecular potential alone does not give the
flexibility required to accurately reproduce the experimental
fractionation ratios \cite{Chialvo2009}.

To determine the reason for the inversion observed in fractionation
above 500 K in both experiment and the q-TIP4P/F model, we decompose
the contributions to the fractionation ratio by noting that the
quantum contribution to the kinetic energy in Eq. \ref{eq:kinetic} is
the dot product of two vectors. The overall kinetic energy is
invariant to the coordinate system used to evaluate it, thus when the
kinetic energy is calculated in the standard Cartesian basis all three
components will average to the same number due to the isotropy of the
liquid. To gain further insight, we instead use the internal
coordinates of the water molecule and determine the contribution to
$10^3 \ln \alpha_{l-v}$ arising from the OH bond vector, a vector in
the plane of the molecule, and the vector perpendicular to the
molecular plane as shown in Fig. \ref{fig:waterbeads}.  This is
similar to the approach taken by Lin {\em et. al.} in the different
context of investigating the proton momentum disitribution in ice
\cite{Lin2011}. The results of this decomposition are shown in Table
\ref{ta:decompose_300} for 300 K where D is experimentally seen to
favor the liquid, and Table \ref{ta:decompose_620} for 620 K where D
is experimentally seen to favor the vapor.

From Table \ref{ta:decompose_300}, we see that all of the models have
largely similar contributions from the two directions orthogonal to
the OH bond and that both are positive and therefore favor the D excess
in the liquid. The contribution perpendicular to the plane is
noticeably larger than the contribution in the plane. As shown in
Eq. \ref{eq:deltaA}, the values depend on the change in the quantum
kinetic energy between H and D in the liquid and the vapor, which in
turn is determined by how much the H or D atom's position uncertainty
is restricted by interacting with the other water molecules in the
liquid or vapor. Since in the vapor there is little confinement in
plane orthogonal to the water, a larger contribution from that
direction is expected; in the liquid, other molecules are present
which restrict expansion in that coordinate.

In all cases the OH contribution is negative, indicating that there is
less confinement in the position of the H atom in that direction in
the liquid than in the vapor because the hydrogen atom participates in
hydrogen bonding allowing the OH chemical bond to stretch more
easily. However, comparing the anharmonic models (Aq-SPCFw and
q-TIP4P/F) with their harmonic counterparts (q-SPCFw and Hq-TIP4P/F),
we observe a 10 fold increase in the values arising from the OH
contribution which gives rise to a larger cancellation of the positive
contributions from the two orthogonal vectors. It is this cancellation
that leads to the much better agreement with the experimental data at 300
K.

We now turn now to Table \ref{ta:decompose_620}, which shows the
contributions to the fractionation ratio at 620 K, a regime where
experimentally D is preferred in the lighter phase. This is closer to
the classical limit where the fractionation would be
zero, thus, as expected, each component is reduced in magnitude
compared to the lower temperature data in Table
\ref{ta:decompose_300}. However, the relative decrease in each
component varies. The contributions arising from the
in-plane and out-of-plane contributions orthogonal to OH decrease by
a factor of 7-8 whereas the OH contribution falls by a factor of
only $\sim$4.  For the anharmonic models, the negative OH
contribution outweighs the positive components in the other two
directions leading to an inversion of the fractionation compared to
that seen at 300 K in agreement with the experimental observation of
-2. The reason the OH component falls off more slowly is that this
direction is dominated by stretching of the OH chemical bond, which is
a high frequency coordinate, so even at high temperatures, quantum
mechanics plays a noticeable role. In contrast, the two directions
orthogonal to the OH direction are lower frequency and so approach the
classical limit more rapidly as the temperature is increased. Thus,
the reason the H/D fractionation is low around 600 K is not due to the
fact that all contributions are individually low but rather that they
nearly exactly cancel at this point due to the different rates at
which the components approach the classical limit.

Finally, we computed the fractionation ratio for the TTM3-F water
model\cite{Fanourgakis2008}, which is known to have a very large
cancellation of its quantum mechanical effects at 300 K
\cite{Habershon2009,Zeidler2011}. This model was fit to {\em ab
  initio} calculations using a potential form incorporating anharmonic
flexibility, geometry dependent charges, and polarizability on the
oxygen site. As such, it represents the current ``gold standard" of
parameterized water models and has been used extensively in recent
studies probing the effects of quantum mechanical fluctuations on
water \cite{Habershon2008,Paesani2011,Liu2011}. To see if the large
cancellation of quantum effects predicted by this model is consistent
with experimental fractionation ratios, we calculated the value at 300
K which yielded a value of $10^3 \ln \alpha_{l-v}=-57$. This is in
qualitative and quantitative disagreement with the experimental
results, since it predicts D to be favored in the vapor at all
temperatures. This model therefore over-predicts competing quantum
effects in water and hence care should be taken  concerning 
its predictions on the effects of quantum fluctuations
on water's structure and dynamics. However, based on our discussion
above it is likely that reparamaterization of the OH bond
anharmonicity could correct this discrepancy.

\section{Conclusion}

In conclusion, we have shown that including anharmonicity in the OH
bond when modeling water is essential to obtain agreement with the
experimentally observed H/D fractionation ratios and that these ratios
provide an excellent method to assess the accuracy of the quantum
effects predicted by models of water. Since it has recently been shown
that the competition between quantum mechanical effects applies to
other hydrogen bonded systems \cite{Li2011}, it is likely that many of
our conclusions will be relevant to understanding isotopic
fractionation in these systems. Additionally, while we only considered
equilibrium fractionation in this work, which can be calculated
exactly for a given potential energy model using PIMD simulations,
many water processes occurring in the world's atmosphere are
non-equilibrium ones. While including the effects of quantum
fluctuations on dynamics is a much more challenging feat, the recent
development of efficient condensed phase quantum dynamics approaches 
\cite{Cao1994,Wang1998,Craig2004} should allow
insights to be gained into kinetic fractionation processes. These
directions will form the basis of future work.

\begin{acknowledgments}
The authors gratefully thank Joseph Morrone and David Selassie for
helpful comments and a critical reading of this
manuscript. This research was supported by a grant to
B.J.B. from the National Science Foundation (NSF-CHE-0910943).
\end{acknowledgments}

\end{article}

\begin{table}
\caption{Decomposition of the components leading to the fractionation ratio, $10^3 \ln \alpha_{l-v}$ at 300 K. 
OH is the contribution to $10^3 \ln \alpha_{l-v}$ arising from quantum kinetic energy down the OH direction, 
plane is that arising from the vector in the molecular plane orthogonal to the OH bond and Orth is the contribution 
down the vector orthogonal to the molecular plane. The type column denotes whether the water model for which the data 
was calculated uses a harmonic or anharmonic description of the OH bond.}
\label{ta:decompose_300}
\begin{tabular*}{0.5\hsize}{@{\extracolsep{\fill}}cccccc}
Model & Stretch & OH & Plane & Orth & Total\cr
\hline
q-SPCFw     &  Harmonic   & -17   & 92 & 141 & 216   \cr
Hq-TIP4P/F  &  Harmonic   & -15   & 84 & 146 & 215   \cr
q-TIP4P/F   &  Anharmonic & -152  & 92 & 164 & 104  \cr
Aq-SPCFw    &  Anharmonic &  -149 & 98 & 151 & 100 \cr 
\hline
\end{tabular*}
\end{table}

\begin{table}
\caption{Decomposition of the components leading to the fractionation ratio, 
$10^3 \ln \alpha_{l-v}$ at 620 K. The labeling is identical to that in Table \ref{ta:decompose_300}.}
\label{ta:decompose_620}
\begin{tabular*}{0.5\hsize}{@{\extracolsep{\fill}}cccccc}
Model & Stretch & OH & Plane & Orth & Total\cr
\hline
q-SPCFw     &  Harmonic   & -3 & 13 & 20 & 30   \cr
Hq-TIP4P/F  &  Harmonic   & -3 & 11 & 20 & 28 \cr
q-TIP4P/F   &  Anharmonic & -40 & 13 & 22 & -5 \cr
Aq-SPCFw    &  Anharmonic &  -40 & 13 & 21 & -6\cr
\hline
\end{tabular*}
\end{table}


\begin{thebibliography}{10}

\bibitem{Hoefs1997}
Hoefs J
\newblock (1997) \emph{Stable Isotope Geochemistry}
\newblock (Springer-Verlag, Berlin).

\bibitem{Worden2007}
Worden J, Noone D, Bowman K
\newblock (2007) Importance of rain evaporation and continental convection in
  the tropical water cycle.
\newblock \emph{Nature} 445:528.

\bibitem{Kuharski1984}
Kuharski RA, Rossky PJ
\newblock (1984) Quantum mechanical contributions to the structure of liquid
  water.
\newblock \emph{Chem. Phys. Lett.} 103:357--362.

\bibitem{Wallqvist1985}
Wallqvist A, Berne BJ
\newblock (1985) Path-integral simulation of pure water.
\newblock \emph{Chem. Phys. Lett.} 117:214--219.

\bibitem{Paesani2006}
Paesani F, {et~al.}
\newblock (2006) An accurate and simple quantum model for liquid water.
\newblock \emph{J. Chem. Phys.} 125:184507--11.

\bibitem{Miller2005}
Miller TF, Manolopoulos DE
\newblock (2005) Quantum diffusion in liquid water from ring polymer molecular
  dynamics.
\newblock \emph{J. Chem. Phys.} 123:154504.

\bibitem{Chen2003}
Chen B, Ivanov I, Klein ML, Parrinello M
\newblock (2003) Hydrogen bonding in water.
\newblock \emph{Phys. Rev. Lett.} 91:215503.

\bibitem{Habershon2009}
Habershon S, Markland TE, Manolopoulos DE
\newblock (2009) Competing quantum effects in the dynamics of a flexible water
  model.
\newblock \emph{J. Chem. Phys.} 131:024501--11.

\bibitem{Kushner1999}
Kushner D, Baker A, Dunstall T
\newblock (1999) Pharmacological uses and perspectives of heavy water and
  deuterated compounds.
\newblock \emph{Can. J. Physiol. Pharmacol.} 77:79--88.

\bibitem{Morrone2008}
Morrone JA, Car R
\newblock (2008) Nuclear quantum effects in water.
\newblock \emph{Phys. Rev. Lett.} 101:017801.

\bibitem{Horita1994}
Horita J, Wesolowski DJ
\newblock (1994) Liquid-vapor fractionation of oxygen and hydrogen isotopes of
  water from the freezing to the critical temperature.
\newblock \emph{Geochimica et Cosmochimica Acta} 58:3425--3427.

\bibitem{Vanicek2007}
Vanicek J, Miller WH
\newblock (2007) Efficient estimators for quantum instanton evaluation of the
  kinetic isotope effects: Application to the intramolecular hydrogen transfer
  in pentadiene.
\newblock \emph{J. Chem. Phys.} 127:114309.

\bibitem{Parrinello1984}
Parrinello M, Rahman A
\newblock (1984) Study of an f center in molten kcl.
\newblock \emph{J. Chem. Phys.} 80:860.

\bibitem{Herman1982}
Herman MF, Bruskin EJ, Berne BJ
\newblock (1982) On path integral monte carlo simulations.
\newblock \emph{J. Chem. Phys.} 76:5150--5155.

\bibitem{berne3_1989}
Cao J, Berne BJ
\newblock (1989) On energy estimators in path integral {M}onte {C}arlo
  simulations: Dependency of accuracy on algorithm.
\newblock \emph{J. Chem. Phys.} 91:6359.

\bibitem{Ceriotti2010}
Ceriotti M, Parrinello M, Markland TE, Manolopoulos DE
\newblock (2010) Efficient stochastic thermostatting of path integral molecular
  dynamics.
\newblock \emph{J. Chem. Phys.} 133:124104.

\bibitem{Markland2008}
Markland TE, Manolopoulos DE
\newblock (2008) An efficient ring polymer contraction scheme for imaginary
  time path integral simulations.
\newblock \emph{J. Chem. Phys.} 129:024105.

\bibitem{Markland2008b}
Markland TE, Manolopoulos DE
\newblock (2008) A refined ring polymer contraction scheme for systems with
  electrostatic interactions.
\newblock \emph{Chem. Phys. Lett.} 464:256 -- 261.

\bibitem{Wagner2002}
Wagner W, Pru$\beta$ A
\newblock (2002) The iapws formulation 1995 for the thermodynamic properties of
  ordinary water substance for general and scientific use.
\newblock \emph{J. Phys. Chem. Ref. Data} 31:387.

\bibitem{Hone2006}
Hone TD, Rossky PJ, Voth GA
\newblock (2006) A comparative study of imaginary time path integral based
  methods for quantum dynamics.
\newblock \emph{The Journal of Chemical Physics} 124:154103.

\bibitem{Habershon2008}
Habershon S, Fanourgakis GS, Manolopoulos DE
\newblock (2008) Comparison of path integral molecular dynamics methods for the
  infrared absorption spectrum of liquid water.
\newblock \emph{J. Chem. Phys.} 129:074501.

\bibitem{Paesani2010}
Paesani F, Voth GA
\newblock (2010) A quantitative assessment of the accuracy of centroid
  molecular dynamics for the calculation of the infrared spectrum of liquid
  water.
\newblock \emph{The Journal of Chemical Physics} 132:014105.

\bibitem{Chialvo2009}
Chialvo AA, Horita J
\newblock (2009) Liquid-vapor equilibrium and isotopic fractionation of water:
  How well can classical water models predict it?
\newblock \emph{J. Chem. Phys.} 130:094509.

\bibitem{Lin2011}
Lin L, Morrone JA, Car R, Parrinello M
\newblock (2011) Momentum distribution, vibrational dynamics, and the potential
  of mean force in ice.
\newblock \emph{Phys. Rev. B} 83:220302--.

\bibitem{Fanourgakis2008}
Fanourgakis GS, Xantheas SS
\newblock (2008) Development of transferable interaction potentials for water.
  v. extension of the flexible, polarizable, thole-type model potential
  (ttm3-f, v. 3.0) to describe the vibrational spectra of water clusters and
  liquid water.
\newblock \emph{J. Chem. Phys.} 128:074506.

\bibitem{Zeidler2011}
Zeidler A, {et~al.}
\newblock (2011) Oxygen as a site specific probe of the structure of water and
  oxide materials.
\newblock \emph{Phys. Rev. Lett.} 107:145501.

\bibitem{Paesani2011}
Paesani F
\newblock (2011) Hydrogen bond dynamics in heavy water studied with quantum
  dynamical simulations.
\newblock \emph{Phys. Chem. Chem. Phys.} 13:19865.

\bibitem{Liu2011}
Liu J, {et~al.}
\newblock (2011) Insights in quantum dynamical effects in the infrared
  spectroscopy of liquid water from a semiclassical study with an ab
  initio-based flexible and polarizable force field.
\newblock \emph{J. Chem. Phys.} 135:244503.

\bibitem{Li2011}
Li XZ, Walker B, Michaelides A
\newblock (2011) Quantum nature of the hydrogen bond.
\newblock \emph{Proc. Nat. Acad. Sci. USA} 108:6369.

\bibitem{Cao1994}
Cao J, Voth
\newblock (1994) The formulation of quantum statistical mechanics based on the
  feynman path centroid density. i. equilibrium properties.
\newblock \emph{J, Chem. Phys.} 100:5093.

\bibitem{Wang1998}
Wang H, Sun X, Miller WH
\newblock (1998) Semiclassical approximations for the calculation of thermal
  rate constants for chemical reactions in complex molecular systems.
\newblock \emph{J, Chem. Phys.} 108:9726.

\bibitem{Craig2004}
Craig IR, Manolopoulos DE
\newblock (2004) Quantum statistics and classical mechanics: Real time
  correlation functions from ring polymer molecular dynamics.
\newblock \emph{J. Chem. Phys.} 121:3368.

\end{thebibliography}
\end{document}